\documentclass[twocolumn,prx,aps,superscriptaddress]{revtex4}
\usepackage{amsmath,amssymb}
\usepackage[colorlinks, citecolor=blue]{hyperref}
\usepackage{amsfonts}
\usepackage{mathrsfs}
\usepackage{graphicx}
\usepackage{float}
\usepackage{amssymb}
\usepackage{color}
\usepackage{dcolumn}
\usepackage{bm}
\usepackage{slashed}
\usepackage{tikz,pstricks}
\usepackage[header,title,page,titletoc]{appendix}
\usepackage{url}

\begin{document}

\title{Critical structure and emergent symmetry of Dirac fermion systems}
\author{Jiang Zhou}
\email{jzhou5@gzu.edu.cn}
\affiliation{Department of Physics, Guizhou University, Guiyang 550025, PR China}

\begin{abstract}
Emergent symmetry in Dirac system means that the system acquires an enlargement of two basic symmetries at some special critical point.
The continuous quantum criticality between the two symmetry broken phases can be described within the framework of Gross-Neveu-Yukawa (GNY) model.
Using the first-order $\epsilon$ expansion in $4-\epsilon$ dimensions, we study the critical structure and emergent symmetry of the chiral GNY model with $N_f$ flavors of four-component Dirac fermions coupled strongly to an $O(N)$ scalar field under a small $O(N)$-symmetry breaking perturbation.
After determining the stable fixed point, we calculate the inverse correlation length exponent and the
anomalous dimensions (bosonic and fermionic) for general $N$ and $N_f$.
Further, we discuss the emergent-symmetry and the emergent supersymmetric critical point for $N\geq4$ on the basis of $O(N)$-GNY model.
It turns out that the chiral emergent-$O(N)$ universality class is physically meaningful if and only if $N<2N_f+4$.
On this premise, the small $O(N)$-symmetry breaking perturbation is always irrelevant in the chiral emergent-$O(N)$ universality class.
Our studies show that the emergent symmetry in Dirac systems has an upper boundary $O(2N_f+3)$, depending on the flavor numbers $N_f$. As a result, the emergent-$O(4)$ and $O(5)$ symmetries are possible to be found in the systems with fermion flavor $N_f=1$,
and the emergent-$O(4)$, $O(5)$, $O(6)$ and $O(7)$ symmetries are expected to be found in the systems with fermion flavor $N_f=2$.
Our result also suggests some rich transitions with emergent-$Z_2\times O(2)\times O(3)$ symmetry and so on.
Interestingly, in the emergent-$O(4)$ universality class, there is a supersymmetric critical point which is expected to be found in the systems with fermion flavor $N_f=1$.
\end{abstract}

\maketitle

\section{Introduction}
In quantum field theory, the Gross-Neveu model is a fundamental model of half spin fermions
and contains a quartic self-interaction of fermion fields\cite{gross,gracey1,handfour}.
It has some interesting properties which are commonly shared by the quantum chromodynamics\cite{gross}.
For example, the phenomena of asymptotic freedom and dynamical symmetry breaking.
Due to its close connections with the phase transitions in graphene,
the general Gross-Neveu theory is also of interest in condensed matter theory.
For instance, it has been suggested that the transition from the semimetal to the Mott-insulating phase in graphene
can be described by what is termed chiral Gross-Neveu model\cite{gracey2,sorella1,herbut1,assaad1,herbut2,janssen1,chan1,knorr1}.
Based on the specific interaction and underlying symmetry,
the chiral Gross-Neveu model is captured by different universality classes,
referring to chiral Ising, XY and $O(3)$ universality classes\cite{zerf1}.
More recently, it is intriguing that the Gross-Neveu model has been found to be related to the AdS/CFT theories\cite{giombi1}.

In the case of chiral Ising universality class,
the Gross-Neveu model is only renormalizable in $1+1$ spacetime dimensions.
Another more widely studied model in this class is the GNY model
\cite{herbut1,assaad1,herbut2,janssen1,chan1,knorr1,zerf1,otsuka1,rosen1,
mihai1,gracey3,lang1,herbut3,roy1,scherer1,torres1,janssen2,torres2,ihrig1},
which is renormalizable in $3+1$ spacetime dimensions.
In contrast to the purely fermionic Gross-Neveu model,
the GNY model can be rewritten by introducing an auxiliary scalar field
so that the Lagrangian is quadratic in fermion field\cite{handfour}.
But now, the scalar field has a canonical kinetic term and the scalar field interacts quartically\cite{zerf1}.
For its wide applications in many areas of physics ranging from high-energy physics to quantum criticality,
the GNY model has been studied by a broad range of different methods,
including functional renormalization group\cite{janssen1,knorr1,gies1}, conformal bootstrap\cite{iliesiu1,iliesiu2,bobev1,bash1} and
$1/N$ expansion\cite{gracey3,mana1,gracey4,gracey5}.
The sign-free quantum Monte Carlo has been broadly used
to study the fermionic quantum criticality of interacting
Dirac fermions on lattice\cite{otsuka1,mchuff1,lzx1,chan1,lzx2,otsuka2,wangl1}.
Finally, the $4-\epsilon$ expansion is a suitable algorithms
to facilitate higher-order calculations for fermionic criticality\cite{rosen1}.
By now, the most accurate calculation
has been carried up to four loops \cite{zerf1,ihrig1}.

In particular, the fermionic criticality in Dirac systems such as graphene
has been extensively discussed\cite{herbut3,otsuka1}.
On graphene's honeycomb lattice, the low-energy gapless Dirac fermions with relativistic dispersion
emerge at the two inequivalent Dirac points in the Brillouin zone.
A typical quantum criticality of the interacting electrons on graphene's lattice
is the transition from a semimetal to an ordered charge-density-wave which breaks sublattice symmetry\cite{herbut1,chan1},
triggered by sufficient large nearest-neighbor repulsion interaction.
Another typical example on this lattice is the semimetal-antiferromagnetic transition
favored by an onsite Hubbard repulsion\cite{otsuka1,assaad1,janssen1}.
Other examples can be constructed according to the ordered partner on the specific lattice
\cite{otsuka3,raghu1,ryu1,hou1,roy2}.
The nature of the quantum phase transition for interacting Dirac fermions
has been under debate for a long time\cite{sore1,herbut1}.
However, recent developments on the fermion-driven quantum critical point (FIQCP)
suggest a second-order continuous phase transition between the semimetallic and the gapped phase
\cite{scherer1,torres1,torres2,janssen2,lzx2,classen1,jsk1,jsk2}.
The gapless Dirac fermions emerge as a new degrees of freedom, consequently,
their quantum criticality can effectively be described by a chiral transitions appearing in
different $2+1$ dimensional Gross-Neveu model, giving rise to the chiral universality classes\cite{zerf1}.
Even in those system which supports cubic terms of order parameter,
the strong fluctuations of gapless fermions would render the putatively first-order transition continuous,
this is right the central idea of FIQCP\cite{classen1,lzx2,yin1,yin2}.
Based on the honeycomb lattice, the evidence for FIQCP have been proposed near
the semimetal-Kekule valence-bond-solid transition\cite{lzx2,classen1}, similar scenario was also proposed in three-dimensional double-Weyl semimetals for semimetal-nodal-nematic order transition\cite{jsk2}.
The FIQCP can be traced back to the fluctuations of gapless Dirac degrees of freedom
at criticality.

Aside from the close connections with the quantum criticality
from a Dirac semimetal to a symmetry broken phase,
the GNY model also has connections with both supersymmetry
\cite{sslee1,sslee2,grover1,jsk3,lzx3,witc1,rahm1}
and emergent symmetry at criticality
\cite{sato1,senthil1,grover2,tanka1,nahum1,lzx5,sree1,roy3}.
It is intriguing that the critical point possess some symmetries
which absent in the original model at the microscopic level.
For instance, right at the fixed point of the special GNY model in which
the velocities for both massless Dirac fermions and relativistic bosons are not equal to each other,
the Lorentz symmetry get restored at criticality, leading to the notation of an emergent Lorentz symmetry\cite{roy4}.
The emergent symmetry at criticality can be attributed to
the presence of fluctuations of new degrees of freedom (gapless Dirac fermions).

In addition, the enlargement of symmetry was put forward to explain the continuous phase transition
between two phases with different broken symmetries
\cite{sato1,senthil1,grover2,tanka1,nahum1,lzx5,sree1,roy5,sandvik1,sandvik2,zhang1,roy3,pujari1}.
It is argued that the deconfined quantum critical point, which separates the Neel and valence-bond-solid orders of
spin-1/2 Heisenberg quantum antiferromagnet on square lattice, possesses an enlarged emergent symmetry
\cite{senthil2,senthil3,mzy1,wangc1}.
Very recently, the similar enlarged emergent symmetry has been stressed
in the Dirac system with competing orders\cite{zhou1,ghae1,janssen2,roy3}.
As pointed above, the critical properties of the transition
from a semimetal to a gapped broken phase can be captured by GNY model\cite{xxy2,liu3,liu4,sei1},
therefore it is believed that the GNY model is accessible to
the emergent symmetry with an appropriate way\cite{classen6,roy6}.
Indeed, on the basis of GNY model, it is stressed that the emergent symmetry is responsible for
various deconfined transitions in Dirac systems\cite{grover1,lzx5,roy3,lzh1}.
On the other hand, the GNY model also connect closely with the supersymmetry
\cite{sslee1,sslee2,grover1,jsk3,lzx3}.
For example, the GNY model with $1/4$ flavor of four-component fermions
in the chiral Ising universality class relates to the $\mathcal{N}=1$ supersymmetry\cite{zerf1},
and the GNY model with $1/2$ flavor of four-component fermions
in the chiral XY universality class relates to the $\mathcal{N}=2$ supersymmetry\cite{lzx3}.

Although the quantum criticality from a semimetal to an $O(N)$ symmetry broken phase for $N\leq3$
has achieved satisfactory understood,
the impact of fermion degrees of freedom
on the stability of the quantum critical point with emergent-$O(N)$ symmetry for $N\geq4$
is under debate\cite{janssen2}.
For the fermionic criticality in Dirac systems,
a variety of critical points with emergent symmetry have been put forward\cite{ryu1,roy3,pujari1,roy5},
by now, however, what kinds of emergent symmetry is allowed is still unclear.
Besides, whether the supersymmetric quantum critical point can emerge
in the chiral emergent-$O(N)$ universality class remains unknown so far.
The stability of the critical point with emergent-$O(N)$ symmetry, as far as we are aware,
is partly answered in Refs.[\onlinecite{janssen2}] and [\onlinecite{roy3}].
Motivated by these issues, in this paper,
without asking the specific lattice model as well as the specific forms of symmetry broken orders ,
we study the critical structure and the emergent symmetry
of the chiral GNY model with $N_f$ flavors four-component fermions coupled strongly to an $O(N)$ scalar field.
The present study has three purposes. The first is to determine the meaningful fixed point that controls
the critical properties in the emergent-$O(N)$ universality class.
The second is to confirm the reasonable emergent-$O(N)$ symmetry in Dirac systems.
Our final purpose is to find the possible supersymmetric critical point in the chiral emergent-$O(N)$ universality class.
To investigate the impact of gapless Dirac fermion degrees of freedom,
we also introduce a small $O(N)$-anisotropy that breaks $O(N)$ symmetry in the chiral GNY model.

The rest of the paper is organized as follows. We define the chiral GNY model in Sec.\ref{section2}.
In Sec.\ref{section3}, we review the basic renormalization group (RG) procedure
and give our results for beta functions and anomalous dimensions.
In Sec.\ref{section4}, we discuss the stability and the emergent symmetry for the emergent-$O(N)$ fixed point,
the supersymmetric quantum critical point
in the chiral emergent-$O(N)$ universality class is also discussed in this section.
Finally, our conclusions and some comments are provided in Sec.\ref{section5}.
More details for the determination of renormalization
constants are presented in Appendix \ref{appendixa}.

\section{The Gross-Neveu-Yukawa model}\label{section2}
We first define the chiral GNY model under investigation.
As pointed in the introduction,
for various interacting relativistic fermions systems in $2+1$ dimensions,
the quantum phase transition
towards a symmetry-broken phase can be captured by the chiral GNY model
in which the fermions couple strongly to a multicomponent boson fields via Yukawa coupling.
Formally, the general chiral $O(N)$-GNY model in $D$-dimensions Euclidean spacetime can be described
by the effective action
\begin{equation}\label{e1}
    S=\int d^Dx(\mathcal{L_{\psi}}+\mathcal{L_{\phi}}+\mathcal{L_{\psi\phi}}).
\end{equation}
The Lagrangian for fermions is simply abbreviated as
\begin{equation}
\mathcal{L}_{\psi}=\bar{\psi}_{i}i\slashed{\partial}\psi_{i},
\end{equation}
where the notation $\slashed{\partial} =\gamma_{\mu }\partial_{\mu }$ is the Feynman slash, and the $4\times 4$ gamma matrices $\gamma_{\mu }$ form a four-dimensional representation of the Clifford algebra, i.e., $\{\gamma_{\mu },\gamma_{\nu }\}=2\delta_{\mu \nu }1_4$, with the indices $\mu $, $\nu =0,1,...,D-1$ and $1_4$ denotes the identity matrix. The Dirac spinor $\psi _{i}$ is a four-component fermion spinor and its conjugate is defined as $\bar{\psi}_{i}={\psi}_{i}^{\dag}\gamma_0$. For generality, we have introduced $N_f$ flavors of four-component spinor such that the Dirac spinor carries a flavor index $i$, $i=1,...,N_f$. The summation convention over repeated indices is also assumed here and in the following.

The second term in (\ref{e1}) describes the purely bosonic part. Explicitly, it takes the form as
\begin{equation}
\mathcal{L}_{\phi}=\frac{1}{2}(\partial_{\mu }\phi _{a})^{2}+\frac{m^2}{2}\phi _{a}\phi _{a}+\frac{\lambda_1 }{4!}(\phi_{a}\phi_{a})^{2}+\frac{\lambda_2}{4!}(\phi_{a})^{4}.
\end{equation}
Here, the index $a$ in $\phi_a$ takes the value range from $1$ to $N$.
In addition to the kinetic term and quartic interactions with strength $\lambda_1$, a small
anisotropy (with strength $\lambda_2$) that breaks $O(N)$ symmetry is added
to investigate the impact of gapless Dirac fermion degrees of freedom.
The mass-square $m^{2}$ plays the role of tuning parameter for phase transition, $m^{2}>0$ corresponds to symmetric phase
with $\langle\phi _{a}\rangle =0$, $m^{2}<0$ corresponds to symmetry broken phase and $m^{2}=0$ at criticality.
In the symmetry broken ground state, the scale field $\phi _{a}$ acquires a nonzero vacuum expectation,
then the fermion mass is generated dynamically.
For the pure $O(N)$ scalar model,
although the symmetry is broken by the small anisotropy presented in the last term,
it is shown that the symmetry is restored at criticality for $N<3$\cite{kleinert1,varn1}.
The $O(N)$ fixed point is unstable under anisotropy for $N>3$.
In this paper, we will demonstrate the impact of gapless Dirac fermion degrees of freedom
on the stability of the $O(N)$ fixed point.

Finally, $\mathcal{L}_{\psi\phi }$ defines the Yukawa coupling between gapless Dirac fermions and $O(N)$-symmetric scalar fields with strength $g$:
\begin{equation}
\mathcal{L}_{\psi \phi }=g\bar{\psi}_{i}\left[(\Sigma_{a})_{4N_f\times4N_f}\cdot\phi_{a}\right]\psi_{i}.
\end{equation}
Each $\Sigma_{a}$ signals the broken pattern of various gapped phases,
their dimension depend on the specific lattice model.
More precisely, the dimension for these sigma matrices coincides with the components of Dirac spinor.
For example, the spinless fermions on honeycomb lattice define a four-component Grassmann spinor,
the dimension for the sigma matrix is four. On the other hand,
the spinful fermions on honeycomb lattice define an eight-component Dirac spinor,
now the dimension for the sigma matrix is eight.

To continue the following calculation, commutating rules between these sigma matrices and Dirac gamma matrices are required.
In the chiral Ising universality class for $N=1$,
the GNY model Eq.(\ref{e1}) includes an one-component real scalar field,
$\Sigma_1$ is a trivial identity matrix in this case,
so we have $[\gamma_{\mu},\Sigma_1]=0$ for chiral Ising-GNY model.
In the chiral XY universality class for $N=2$, the GNY model includes a complex order parameter, now the Yukawa term
can be generally written as
$\mathcal{L}_{\psi \phi }=g\bar{\psi}_{i}(\phi_{1}+i\gamma^5\phi_2)\psi_{i}$,
where $\{\gamma^5,\gamma_{\mu}\}=0$ and the explicit choice of $\gamma^5$ depends on the specific model\cite{roy1,zerf3,janssen4}.
Note that $\gamma^5S_{\psi}(p)=-S_{\psi}(p)\gamma^5$, where $S_{\psi}(p)$ is the fermion propagator.
Since in any non-vanished Feynman loop,
each fermion propagator is attached by two $i\gamma^5$ factors at both sides respectively,
then the minus generated from the interchange of $\gamma^5$ and $S_{\psi}(p)$ is countered by $i^2=-1$.
Therefore, the vanished commutator $[\vec{\Sigma}_{\text{XY}},\gamma_{\mu}]=0$ works in the practical calculations
for chiral XY-GNY model.
Finally, in the chiral $O(3)$ universality class for $N=3$,
the three-component order parameters break spin-rotational symmetry spontaneously.
An explicit choice of the Yukawa interactions is given in Ref.[\onlinecite{janssen1}],
and the commutator $[\vec{\Sigma}_{O(3)},\gamma_{\mu}]=0$ is satisfied.

Since the emergent $O(N)$ symmetry for $N\geq4$ can be builded on
the competing orders that break these three basic symmetries\cite{grover2,roy3,ryu1},
it is reasonable to assume the commutating rules
\begin{equation}\label{comm}
  [\Sigma_a,\gamma_{\mu}]=0, \quad \forall a, \mu,
\end{equation}
in chiral $O(N)$-GNY model.
This vanished commutator is essential for the derivation of fixed points and critical exponents in
the chiral emergent-$O(N)$ universality class.
In the practical calculation, all beta and eta functions are independent from the explicit matrix
representation of the Dirac gamma matrix, and only the Clifford algebra and the dimensions
of the representation matrix are required in the renormalization group calculations.
At the tree level, the scaling dimensions for the field variables
and coupling constants can be identified from Eq.(\ref{e1}),
\begin{align}
[\psi]=\frac{D-1}{2}, \quad [\phi]=\frac{D-2}{2},\\
[\lambda]=4-D, \quad [g]=\frac{4-D}{2}.
\end{align}
The Yukawa coupling $g$, quartic $O(N)$-interaction $\lambda_1$,
as well as the strength for small $O(N)$-symmetry-breaking perturbations $\lambda_2$,
are all marginal at the upper critical dimension $D_{uc}=4$, implying that the critical properties
is accessible via standard epsilon expansion in $D=4-\epsilon$ spacetime dimensions.

\section{Field theory and Renormalization group}\label{section3}
This section present the renormalization group (RG) analysis of the chiral $O(N)$-GNY model
under a small $O(N)$-symmetry-breaking perturbation, see Eq.(\ref{e1}).
To perform standard RG analysis in $4-\epsilon$ dimensions,
we employ dimensional regularization and modified minimal subtraction (MS) scheme.
The bare Lagrangian is defined by replacing the field variables and different couplings with their bare counterparts,
i.e.,
\begin{equation}
\psi\rightarrow \psi_0, \phi\rightarrow \phi_0, \lambda\rightarrow \lambda_0, g\rightarrow g_0, m^2\rightarrow m^2_0.
\end{equation}
The renormalized Lagrangian is then written as
\begin{align}\label{rl}
\mathcal{L}&=Z_{\psi}\bar{\psi}_ii\slashed{\partial}\psi_i+\frac{1}2Z_{\phi}(\partial_{\mu}\phi _{a})^2+
             \frac{1}{2}Z_{\phi }Z_{m^2}m^2\phi_a^{2} \notag \\
            &+\frac{\lambda_1}{4!}\mu^{\epsilon }Z_{\lambda_1}Z_{\phi }^{2}\left(\phi_a\phi_{a}\right)^2
             +\frac{\lambda_2}{4!}\mu^{\epsilon }Z_{\lambda_2}Z_{\phi }^{2}\left(\phi_a\right)^4  \notag \\
            &+g\mu^{\epsilon /2}Z_g Z_{\psi}\sqrt{Z_{\phi}}\bar{\psi}_i\left(\Sigma_a\phi _a\right)\psi_i.
\end{align}
where $\mu $ is the energy-scale parameterizing the RG flow of the coupling constants.
These different $Z$-factors are named renormalization constants,
which are used to absorbed the divergence in the loop corrections.
The wave-function renormalization constants $Z_{\psi}$, and $Z_{\phi}$,
relate the bare and renormalized field variables upon the field rescalings
$\psi _{0}=\sqrt{Z_{\psi }}\psi $, $\phi _{0}=\sqrt{Z_{\phi }}\phi$.
Accordingly, the bare mass-square, quartic couplings,
and the Yukawa coupling are related to their dimensionless partner as following,
\begin{gather}
          m^2_0=\mu^{2} m^2 Z_{m^2}, \\
   \lambda_{10}=\mu^{\epsilon}\lambda_1 Z_{\lambda_1},\label{g1} \\
   \lambda_{20}=\mu^{\epsilon}\lambda_2 Z_{\lambda_2},\\
          g_{0}=\mu^{\epsilon/2}g Z_{g}.\label{g0}
\end{gather}
We have also introduced the rescaling couplings
\begin{equation}
  g\rightarrow g\mu ^{\epsilon /2}, \lambda_1 \rightarrow \lambda_1 \mu^{\epsilon }, \lambda_2 \rightarrow \lambda_2 \mu^{\epsilon },
\end{equation}
such that different couplings in the renormalized Lagrangian are dimensionless,
this lead to explicit energy-scale dependencies Lagrangian.

The RG beta functions are defined as the logarithmic derivatives with respect to $\mu$,
$\beta(X)=dX/d\ln\mu$, where $X=m^2,\lambda_1,\lambda_2,g$.
These beta functions can be derived using the fact that the bare value are independent of $\mu$.
Defining the mass-square anomalous dimensions $\gamma_{m^2}=d\ln Z_{m^2}/d\ln\mu$, we find
\begin{equation}
   \beta(m^2)=-(2+\gamma_{m^2})m^2.
\end{equation}
The $\ln\mu$ derivative of Eqs. (\ref{g1})-(\ref{g0}) give the beta function of dimensionless couplings:
\begin{gather}
   \beta(\lambda_1)=-\epsilon\lambda_1-\frac{1}{Z_{\lambda_1}}\frac{\partial Z_{\lambda_1}}{\partial \ln \mu}\lambda_1,\label{b1}\\
   \beta(\lambda_2)=-\epsilon\lambda_2-\frac{1}{Z_{\lambda_2}}\frac{\partial Z_{\lambda_2}}{\partial \ln \mu}\lambda_2,\label{b2}\\
   \beta(g^2)=-\epsilon g^2-\frac{2}{Z_{g}}\frac{\partial Z_{g}}{\partial \ln \mu}g^2.\label{bg}
\end{gather}
Here, we have introduced the squared Yukawa coupling $g^2$ for notational simplicity.
In terms of these RG beta functions, the inverse correlation length exponent $\nu^{-1}$ is determined by \cite{mihai1,zerf1}
\begin{equation}\label{mu}
    \nu^{-1}=-\frac{d\beta(m^2)}{m^2}=2+\gamma_{m^2}.
\end{equation}
Furthermore, the fermion anomalous dimensions $\gamma_{\psi}$, and boson anomalous dimensions $\gamma_{\phi}$ are obtained from
\begin{equation}\label{andim}
\gamma_{\phi}=\frac{1}{Z_{\phi}}\frac{\partial Z_{\phi}}{\partial \ln \mu}, \quad
\gamma_{\psi}=\frac{1}{Z_{\psi}}\frac{\partial Z_{\psi}}{\partial \ln \mu}.
\end{equation}
Both the beta functions and anomalous dimensions can be determined from the
renormalization constants in the context of the standard perturbative RG.
In the next subsection, we will determine the renormalization constants at the leading order.

\subsection{Renormalization constants}
\begin{figure}[t]
\centering
\includegraphics[width=0.5\textwidth]{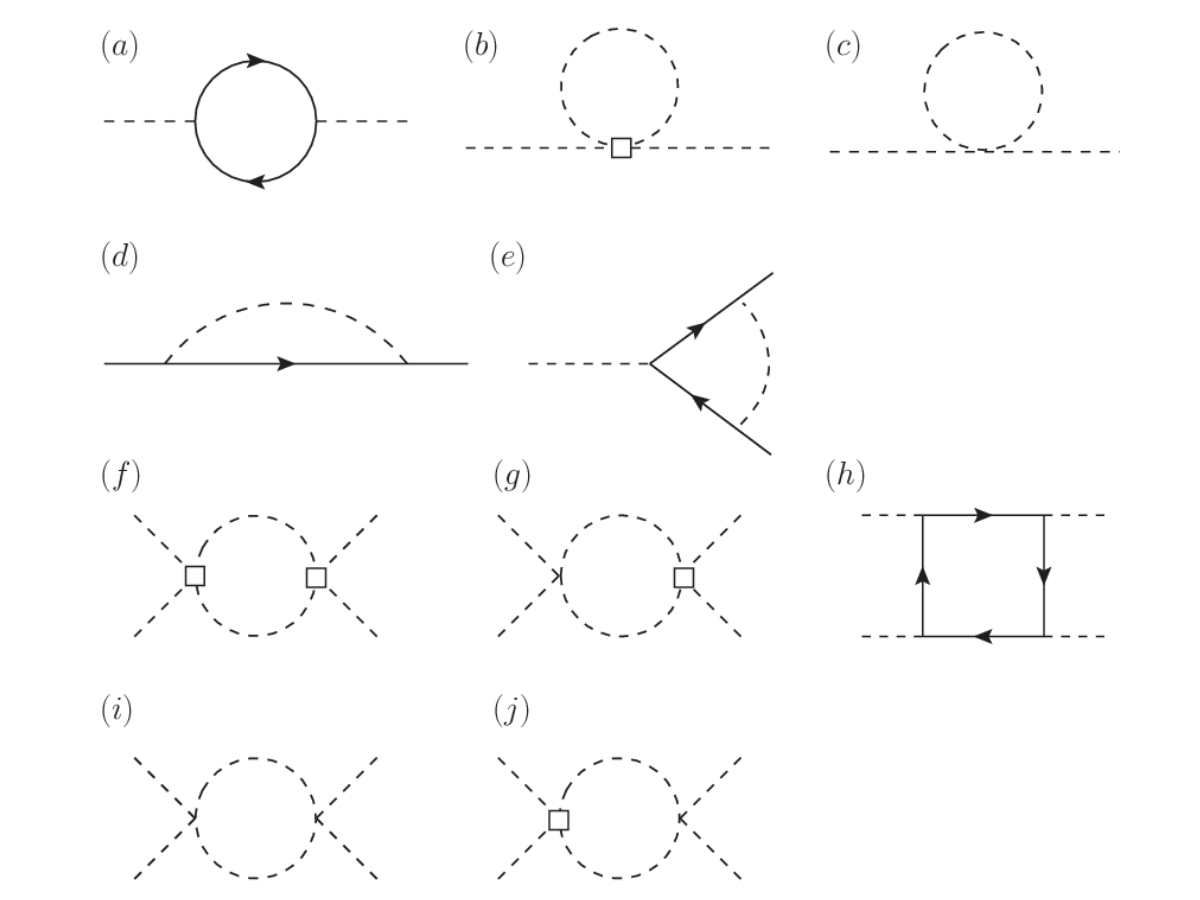}
\caption{One-loop 1PI diagrams. (a), (b) and (c) for boson self-energy, (d) for fermion self-energy, (e) for Yukawa vertex, (f), (g) and (h) for $O(N)$-symmetry quartic vertex, (i) and (j) for $O(N)$-anisotropy quartic vertex.}
\label{fig1}
\end{figure}
In order to access the properties of critical point,
one must calculate all the renormalization constants.
Using dimensional regularization and MS scheme, they depend only on the dimensionless couplings and can be
expanded into the following formal Laurent series\cite{peskin1},
\begin{equation}
Z_X(1/\epsilon,\lambda_1,\lambda_2,g)=1+\sum_{k=1}^{\infty}Z^{(k)}_X(\lambda_1,\lambda_2,g)\frac{1}{\epsilon^k},
\end{equation}
which can be determined order by order in pertubative RG. To the leading order, we expand $Z_X=1+\delta_X$
and demand these $\delta_X$s cancel the ultraviolet divergences of the one-loop corrections.
The one-particle irreducible (1PI) diagrams for two-point boson self-energy,
two-point fermion self-energy,
three-point Yukawa vertex,
$O(N)$-symmetry quartic vertex,
and $O(N)$-anisotropy quartic vertex are indicated in Fig.\ref{fig1}.
To calculate these one-loop 1PI divergent diagrams, the fermion and boson propagators are
\begin{gather}
S^{\psi}_{ab}(p)=-i\delta_{ab}\frac{\slashed{p}}{p^{2}},\\
D^{\phi}_{ij}(p)= \frac{\delta_{ij}}{p^2+m^2}.
\end{gather}
And the complete Feynman rules are illustrated in Fig.\ref{fig2}.
We present a more detailed calculations of these one-loop 1PI diagrams in Appendix \ref{appendixa},
here we only quote the final results:
\begin{gather}
Z_{\psi}=1-Ng^2\frac{K}{\epsilon},\\
Z_{\phi}=1-4N_fg^2\frac{K}{\epsilon},\\
Z_{m^2}=1+(\frac{N+2}{3}\lambda_1+\lambda_2+4N_fg^2)\frac{K}{\epsilon},
\end{gather}
\begin{gather}
Z_{\lambda_1}=1+(\frac{N+8}{3}\lambda_1+2\lambda_2-48N_fg^4/\lambda_1+8N_fg^2)\frac{K}{\epsilon},\\
Z_{\lambda_2}=1+(3\lambda_2+4\lambda_1+8N_fg^2)\frac{K}{\epsilon},\\
Z_{g}=1+(2N_f+4-N)g^2\frac{K}{\epsilon}.
\end{gather}
It is seen from above that all renormalization constants have a simple pole at $\epsilon$.
Here and in the following, the constant $K=1/(4\pi)^2$.
To derive these renormalization $Z$-factors,
we have made use of the sigma matrix with dimensions four but without asking its explicit representation.
The dimensions-dependence on the sigma matrix is generalized by introducing $N_f$ flavors of four-component fermions.
\begin{figure}[t]
\centering
\includegraphics[width=0.5\textwidth]{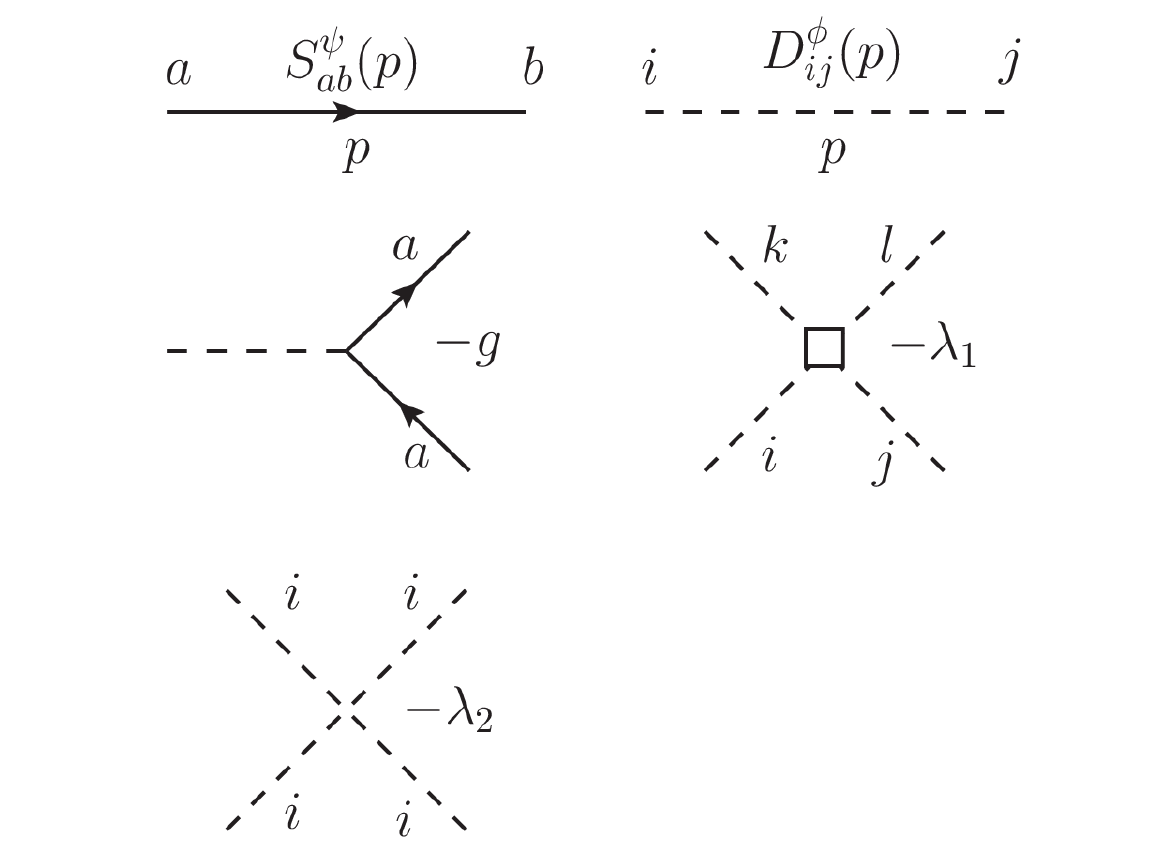}
\caption{Feynman rules for the fermion and boson propagators, Yukawa vertex, $O(N)$-symmetry and $O(N)$-anisotropy quartic vertex.}
\label{fig2}
\end{figure}
\subsection{Beta functions and anomalous dimensions}
To derive the exact expressions of the beta functions, we first use the chain rule, to write
\begin{equation}\label{e30}
     \frac{\partial Z_{X}}{\partial\ln\mu}=\sum_Y\frac{\partial Z_{X}}{\partial Y}\beta(Y),
\end{equation}
where $X,Y\in\{m^2,\lambda_1,\lambda_2,g\}$. Inserting the renormalization $Z$-factors
into Eqs.(\ref{b1})-(\ref{bg}) then produces
the beta functions for $O(N)$ scalar coupling $\lambda_1$,
$O(N)$-anisotropy strength $\lambda_2$, and squared Yukawa coupling $g^2$, respectively.
We find
\begin{align}\label{beta1}
    \beta(\lambda_1)&=-\epsilon\lambda_1+(N+8)K\lambda^2_1/3+2K\lambda_1\lambda_2 \notag\\
                    &+8KN_fg^2\lambda_1-48KN_fg^4,
\end{align}
\begin{align}\label{beta2}
    \beta(\lambda_2)=-\epsilon\lambda_2+3K\lambda^2_2+4K\lambda_1\lambda_2+8KN_fg^2\lambda_2,
\end{align}
\begin{align}\label{betag}
    \beta(g^2)=-\epsilon g^2+(4N_f+8-2N)Kg^4.
\end{align}
Our beta function can be verified in some specific limits.
For example, setting $g=0$, we recover the one-loop beta function
for the $O(N)$ scalar model under a cubic-anisotropy\cite{amit1}.
Setting $g=0$ and $\lambda_2=0$, $\beta(\lambda_1)$ agrees exactly with the results of $O(N)$ scalar model.
In the chiral GNY limit upon setting $\lambda_2=0$,
rescaling the coupling $\alpha/8\pi^{2}\mapsto\alpha$
for $\alpha=g^{2}$, $\lambda_1$, $\lambda_2$,
and replacing the boson self-interactions according to $\lambda/4!\mapsto\lambda$,
our beta function agrees fully with the corresponding expressions derived in the chiral GNY model \cite{zerf1}.

In terms of Eq.(\ref{e30}), the inverse correlation length exponent and
the anomalous dimensions [see Eqs.(\ref{mu})-(\ref{andim})] can be obtained,
to the one-loop order, as
\begin{align}
\nu^{-1}&=2-(\frac{N+2}{3}\lambda_1+\lambda_2+4N_fg^2)K. \label{innu}\\
\gamma_{\phi}&=4KN_{f}g^{2}, \quad \gamma_{\psi}=KNg^2. \label{anphi}
\end{align}
The couplings dependence of these exponents is meant to be evaluated at the stable fixed point that controls the critical behaviors.

\section{RG analysis}\label{section4}
We began the RG analysis by searching for the fixed points of the beta function.
The beta function for the Yukawa coupling exhibits two fixed points: zero and nonzero.
It is easy to see that only the fixed point with finite Yukawa value is stable,
since at which the slope $\partial\beta(g^2)/\partial g^2$ is positive.
In the case of $g^2=0$, the beta functions in Eqs.(\ref{beta1})-(\ref{beta2})
admit four well understood fixed points\cite{varn1,amit1}:
the trivial Gaussian fixed point $(0,0)$,
Ising fixed point $[\epsilon/(3K),0]$,
anisotropic fixed point $[\epsilon/(NK),(N-4)\epsilon/(3NK)]$,
and isotropic Wilson-Fisher fixed point $[3\epsilon/((N+8)K),0]$.
The most intriguing fixed points are the Wilson-Fisher and the anisotropic fixed points as they exchange their stability
at the critical dimensionality $N_c\approx3$, see for example in Ref.[\onlinecite{varn1}].

At the fixed point with finite Yukawa coupling,
solving the common zero of Eqs.(\ref{beta1}) and (\ref{beta2}),
we find four different fixed points:
two Yukawa-Wilson-Fisher (YWF) fixed points
and two fixed points with small $O(N)$-symmetry-breaking anisotropy which we will term anisotropic fixed point (AFP).
Among all those fixed points, we are interesting in the ones with positive $\lambda_1$,
so-called YWF2 fixed point and AFP2, as their stability depend on the value of $N$ and fermion flavors $N_f$.
The YWF2 fixed point locates at
\begin{align}
\lambda_{1*}&=\frac{3}{2}\frac{F+N^{2}-8N-4N_{f}^{2}+16}{(N+8)(2N_{f}-N+4)^{2}}\frac{\epsilon}{K},\\
\lambda_{2*}&=0,\\
        g^2_{*}&=\frac{1}{2(2N_f-N+4)}\frac{\epsilon}{K},
\end{align}
defining
\begin{equation}
F\equiv\sqrt{[(N-4)^{2}+4N_{f}^{2}+20N_{f}N+112N_{f}]F_0},
\end{equation}
with $F_0=(2N_{f}-N+4)^{2}$, and the AFP2 locates at
\begin{align}
\lambda_{1*}&=\frac{W+2N_{f}+N-4}{2N(N-2N_f-4)}\frac{\epsilon}{K},\label{afp1}\\
\lambda_{2*}&=\frac{-2W+(N+2N_f-6)N-4N_f+8}{3N(N-2N_f-4)}\frac{\epsilon}{K},\label{afp2}\\
        g^2_*&=\frac{1}{2(2N_f-N+4)}\frac{\epsilon}{K},\label{afpg}
\end{align}
where $ W=[4N_f^2+4N_f(37N-4)+(N-4)^2]^{1/2}$.
The stable fixed point occurs at the nonzero Yukawa coupling,
thus the chiral $O(N)$ universality class
has strongly boson-fermion coupled critical fluctuations.
We will analysis the interplay between
the boson-fermion coupled critical fluctuations
and the small $O(N)$-symmetry-breaking anisotropy.

\subsection{Stability analysis}
\begin{figure}[t]
\centering
\includegraphics[width=0.48\textwidth]{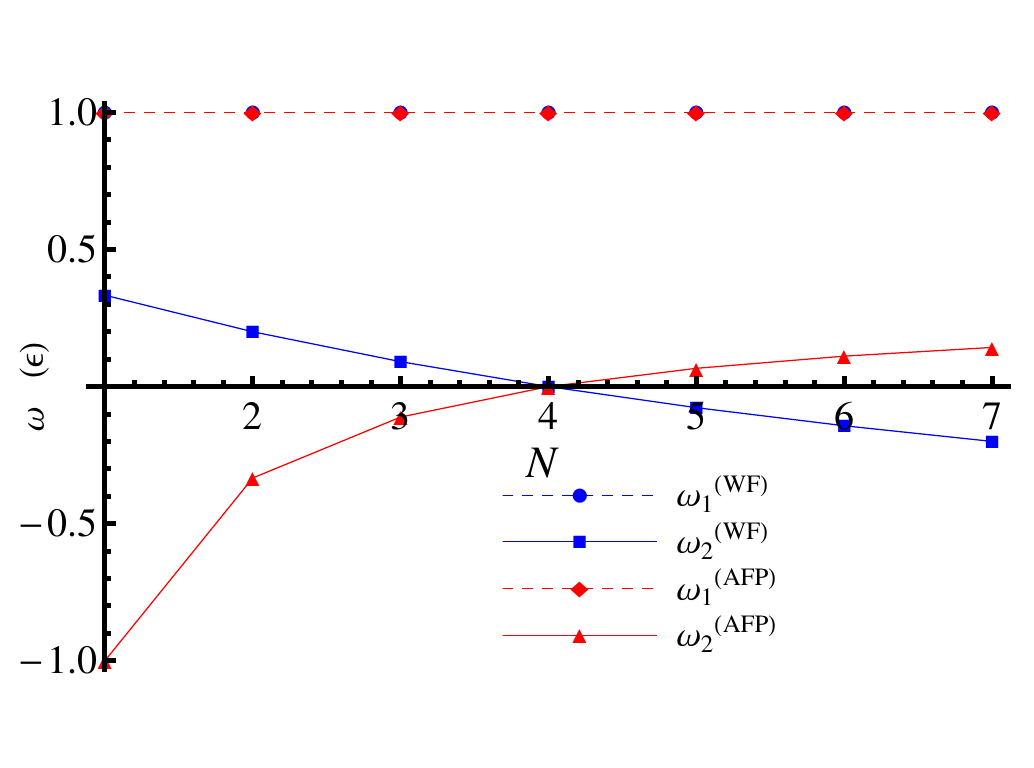}
\caption{Stability exponents as a function of $N$ in the non-Yukawa limit.
$\omega^{\text{(WF)}}$ (blue) and $\omega^{\text{(AFP)}}$ (red) denote the stability exponents evaluated at the isotropic Wilson-Fisher and the anisotropic fixed point, respectively.}
\label{fig3}
\end{figure}
The stability of the fixed points is determined by a matrix $M_{ij}$, which is termed stability matrix and defined as the first derivatives of the beta functions with respect to the couplings
\begin{align}
    M_{ij}=\frac{\partial\beta(\lambda_i)}{\partial\lambda_j},
\end{align}
where $i,j=1,2,3$ and $\lambda_3=g^2$.
The reliable conclusion about the stability of fixed point can be given by calculating the eigenvalues of the stability matrix taken
at the fixed point.
If the real part of the eigenvalues are all positive,
the fixed point is stable and corresponds to a sink.
On the other hand, if the real part of the eigenvalues have opposite sign, the fixed point is of a saddle-point.
A saddle-point type fixed point acquires at least an unstable direction on the surface spanned by the coupling constants.
An important property of the stability matrix is that its $i$-th eigenvalue (denoted by $\omega_{\lambda_i}$)
controls the RG flow approaching the fixed point along $\lambda_i$-direction.
In turn, these eigenvalues are called stability exponents.

For the non-Yukawa limit with $g^2=0$, the stability matrix reads
\begin{align*}
    M_{ij}=\left(\begin{array}{cc}
                   -\epsilon+\frac{N+8}{3}2K\lambda_1+2K\lambda_2 & 2K\lambda_1 \\
                   4K\lambda_2 & -\epsilon+6K\lambda_2+4K\lambda_1
                 \end{array}
    \right).
\end{align*}
To obtain the stability exponents of the fixed point, we calculate the eigenvalues of the stability matrix.
The stability exponents at the isotropic Wilson-Fisher (or named Heisenberg) fixed point
and the anisotropic fixed point are plotted in Fig.\ref{fig3}, respectively.
It is seen from Fig.\ref{fig3} that the critical value $N_c=4$ separates two distinct regimes of the stable fixed point.
For $N<N_c$, the isotropic Wilson-Fisher fixed point is stable.
While for $N>N_c$, the anisotropic fixed point is stable
as the stability exponents $\omega^{(\text{AFP})}_1>0$ and $\omega^{(\text{AFP})}_2>0$ in this regime.
Therefore, both fixed points
merge into a single point and exchange their stability at $N_c$.
For the most accurate value of $N_c$,
early four and five-loop approximations suggest it lies bellow $3$\cite{kleinert1,varn1}.

\begin{figure}[t]
\centering
\includegraphics[width=0.47\textwidth]{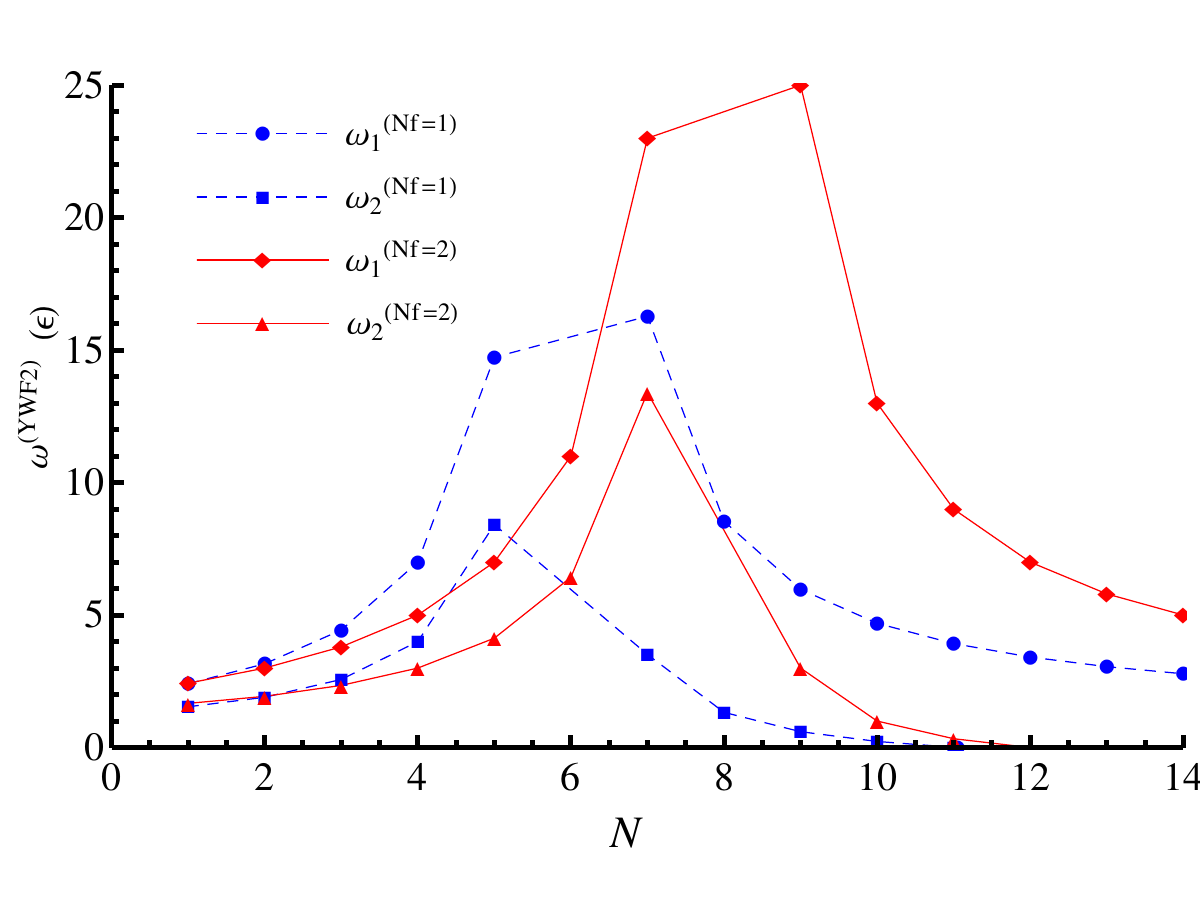}
\caption{Numerical results for the first two stability exponents at the YWF2 fixed point as a function of $N$.
dashed blue line is plotted for $N_f=1$ and red line is plotted for $N_f=2$.}
\label{fig4}
\end{figure}

\begin{figure}[t]
\centering
\includegraphics[width=0.47\textwidth]{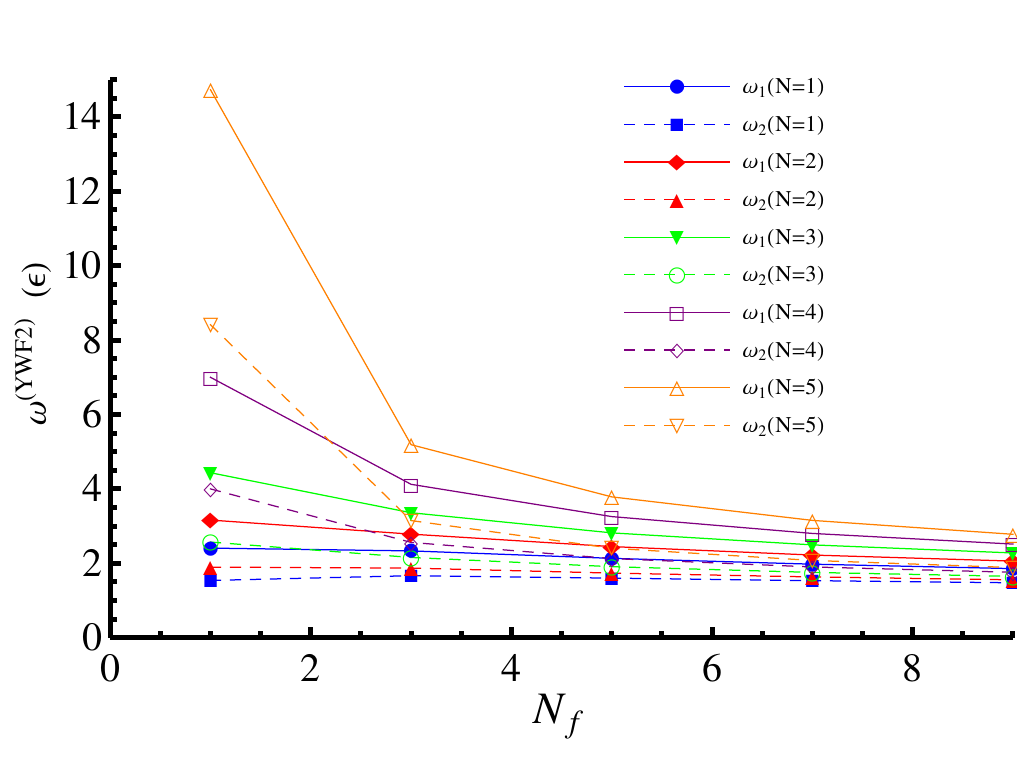}
\caption{Stability exponents of the YWF2 fixed point for different $N$ as a function of $N_f$.}
\label{fig5}
\end{figure}

Let us now concentrate our attentions on the YWF2 fixed point.
In the presence of finite Yukawa coupling,
the stability matrix can be derived from Eqs.(\ref{beta1})-(\ref{betag}).
Diagonalization of the stability matrix
shows that the stability of the YWF2 fixed point depends on $N$.
In Fig.\ref{fig4}, we plot the stability exponents of the YWF2 fixed point for $N_f=1$ and $N_f=2$.
In the case of $N_f=1$, the first exponents are constantly positive for different $N$,
while the second exponent changes its sign at $N\approx11$, see the blue-square in Fig.\ref{fig4}.
This implies that the YWF2 fixed point is stable
and governs the critical behaviors for $N<11$.
In the case of $N_f=2$, the stability exponents are illustrated by red line in Fig.\ref{fig4},
the results are qualitatively similar as that for $N_f=1$ but the stability is separated by $N=12$ (see red triangle).
Further, we also determine the stability of YWF2 fixed point
as the function of flavors of fermions, and the results are shown in Fig.\ref{fig5}.
From Fig.\ref{fig5}, we see that the exponents for different $N$ do not change their sign as $N_f$ increases.
In particular, the YWF2 fixed point in chiral emergent-$O(4)$ and $O(5)$ universality class is stable,
this implies fermion-induced symmetry enhancement in interacting Dirac fermion systems\cite{janssen2,roy3}.

The stable YWF2 fixed point means that the $O(N)$-anisotropy is irrelevant.
Indeed, for $N=1$, the initial model possesses an
exchange symmetry $\lambda_1 \Leftrightarrow\lambda_2$.
As a result, the beta functions should obey the relation
$\beta(\lambda_1+\lambda_2)=\beta(\lambda_1)+\beta(\lambda_2)$,
the anisotropic fixed point and the isotropic fixed point merge to form a new Yukawa Wilson-Fisher (stable) fixed point.
Then the initial model reduce to the chiral Ising GNY model with an effective coupling
$\lambda_{e}=\lambda_1+\lambda_2$ in this special case.


\subsection{RG flows}
Aside from the stability analysis for the fixed point, RG flows provide us with
an alternative way to investigate the properties of the fixed point.
Since the beta function for Yukawa coupling is independent of $\lambda_1$ and $\lambda_2$ [see Eq.(\ref{betag})],
the stable fixed point shares the common Yukawa value.
Therefore, it is sufficient only to plot the projected RG flows in the $\lambda_1$-$\lambda_2$ plane at the finite Yukawa coupling.

\begin{figure*}[t]
\centering
\includegraphics[width=0.95\textwidth]{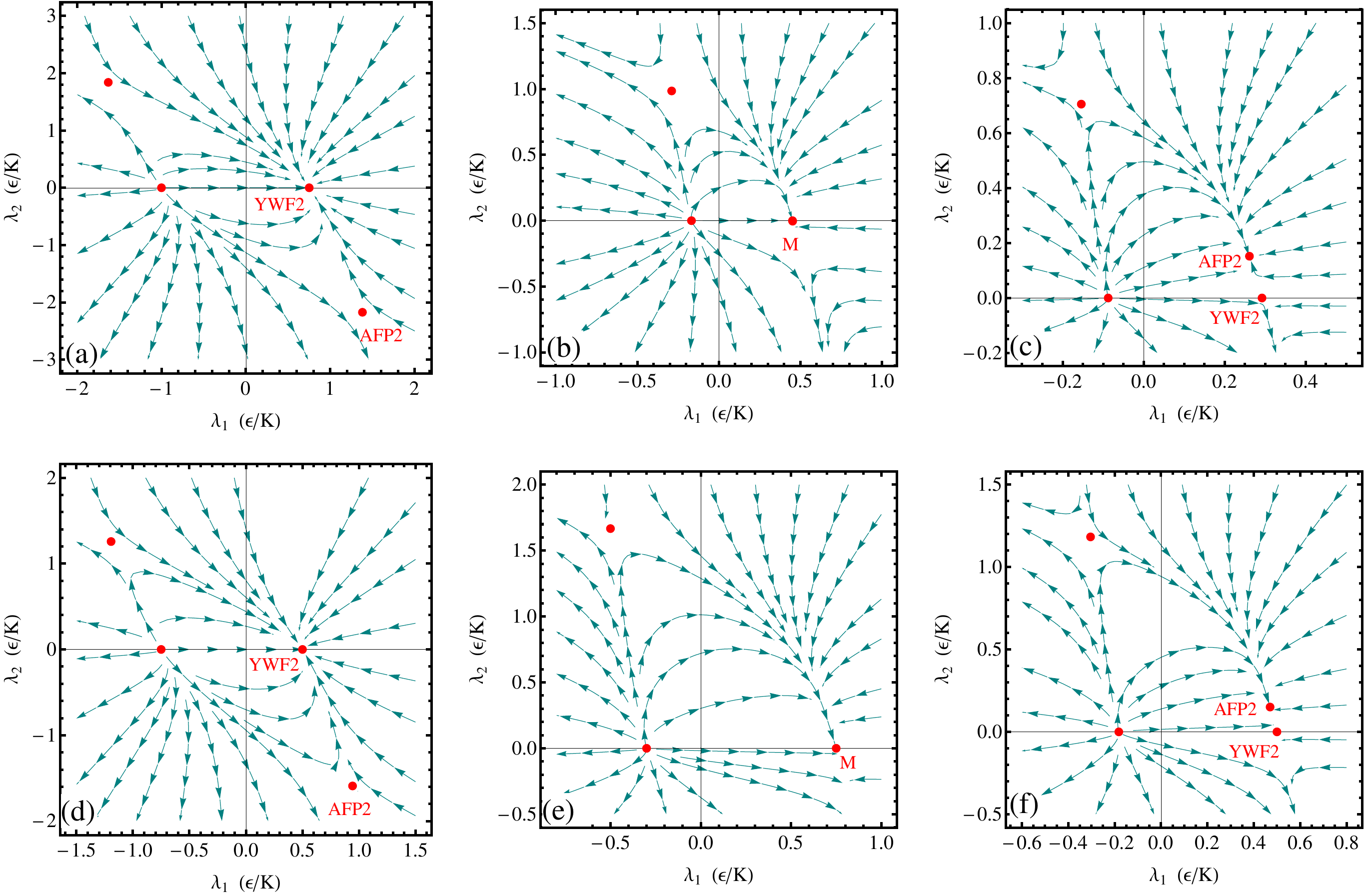}
\caption{Projected RG flows in the $\lambda_1$-$\lambda_2$ plane. The first row (a)-(c) display the flows for $N_f=1$.
Panel (a) is plotted for $N=4$, the YWF2 fixed point is stable and AFP2 is unstable.
With the increase in $N$, two fixed points merge into a single point M at $N\approx11$ in (b).
Panel (c) is plotted for $N=14$, the YWF2 fixed point is unstable and AFP2 is stable in (c).
The second row (d)-(f) display the flows for $N_f=2$.
Panel (d) is plotted for $N=4$, the YWF2 fixed point is stable and AFP2 is unstable.
With the increase in $N$, two fixed points merge into a single point M at $N=12$ in (e).
Panel (f) is plotted for $N=14$, the YWF2 fixed point is unstable and AFP2 is stable in (f).}
\label{fig6}
\end{figure*}

We plot the RG flows for $N_f=1$ in Fig.\ref{fig6}(a)-(c).
For the chiral $O(4)$-GNY model [Fig.\ref{fig6}(a)],
the YWF2 fixed point is a stable point, while the AFP2 is a saddle point.
With the increase in $N$, AFP2 moves gradually toward YWF2 fixed point,
then they merge into a single point M at about $N=11$,
as shown in Fig.\ref{fig6}(b).
Right at the point M, the beta functions have a marginal component in $\lambda_2$-direction.
With further increase of $N$, the YWF2 fixed point acquires one unstable direction,
and the AFP2 turns into a stable one in Fig.\ref{fig6}(c).
Further, the RG flows for $N_f=2$ are illustrated in Fig.\ref{fig6}(d)-(f).
As in the case of $N_f=1$, the YWF2 fixed point is stable and
controls the critical behavior in the chiral $O(4)$ universality class.
With the increase in $N$, the YWF2 fixed point and AFP2 merge into a single point M at $N=12$.
The YWF2 fixed point is stable for $N<12$, while the AFP2 is stable for $N>12$.
These RG flows are in consistent with the stability analysis.

Although the RG flows suggest another stable fixed point (say AFP2) for sufficiently large $N$,
what has to be emphasized is that such a new fixed point is not physically meaningful and cannot be reached in Dirac systems
since the Yukawa coupling now is immeasurable for sufficiently large $N$[see Eq.(\ref{afpg})].
The non-negativity of the squared Yukawa coupling implies that $N$ has an upper boundary.
As a consequence, the emergent symmetry has a maximum value in different Dirac fermion systems.

\subsection{Critical exponents and emergent symmetries}
When the system is tuned to criticality, all couplings flow to the infrared stable
fixed point at which the system exhibits scale invariance.
Close enough to the scale invariant point,
the correlation length and two-point correlation functions have the form of simple power-law.
These power exponents define the critical exponents.
Here, we calculate the inverse correlation length exponent,
bosonic anomalous dimensions,
and the fermionic anomalous dimensions for the chiral $O(N)$-GNY model
at the physically reasonable fixed point.
Furthermore, we also discuss the supersymmetric quantum critical point in the chiral emergent-$O(N)$ universality class.

Near the stable YWF2 fixed point, the anomalous dimensions for bosons as well as fermions, to the leading order in $\epsilon=4-D$, read
\begin{align}
(\eta_{\phi},\eta_{\psi})=\left[\frac{2N_f}{(2N_f-N+4)},\frac{N}{2(2N_f-N+4)}\right]\epsilon \label{abb},
\end{align}
which control the scaling of two-point boson and fermi correlation functions.
Inserting the YWF2 fixed point into Eq.(\ref{innu}), we find for the inverse correlation length exponent
\begin{align}
    \frac1{\nu_y}=2-&\frac{(N+2)(F+N^2-8N-4N_f^2+16)}{2(N+8)(2N_f-N+4)^2}\epsilon \notag\\
                 -&\frac{2N_f}{2N_f-N+4}\epsilon.
\end{align}
For $N_f=2$, the numerical evaluation of the expression provides the following series in $\epsilon$ :
$\nu_y^{-1}=2-0.9524\epsilon$ $(N=1)$ and $\nu_y^{-1}=2-1.2\epsilon$ $(N=2)$,
in full agreement with the previous study in the chiral Ising and chiral XY universality class\cite{zerf1}.
When extrapolated to $\epsilon=1$, the anomalous dimensions [see Eq.(\ref{abb})]
at the linear order in $\epsilon$ must be non-negative.
Then, the non-negativity of the anomalous dimensions imposes a constraint on $N$,
or on the emergent $O(N)$ symmetry, reading
\begin{align}\label{cont}
    N<2N_f+4.
\end{align}
Meanwhile, we note that this constraint is also a natural result of the measurability of Yukawa coupling.
Due to the constraint, the AFP2 is therefore a non-physical fixed point
for $N_f=1$ and $N_f=2$.

The numerical series for the inverse correlation length exponent $\nu_y^{-1}$
have been provided in Table \ref{tab1}.
In general, the chiral $O(N)$ universality class are divided into two classes.
The first class is the conventional chiral $O(N)$ universality class
in which the ordered phase breaks some basic symmetries.
Another class is the chiral emergent-$O(N)$ universality class
in which the critical point possesses an enlargement of the basic symmetries.
For instance, the chiral Ising-GNY model is relevant to
the quantum criticality of semimetal-CDW transition on graphene's honeycomb lattice,
where the ordered state breaks $Z_2$ sublattice symmetry spontaneously.
The chiral XY-GNY model has connections with superconducting or Kekul\'{e} valence-bond-solid transition in graphene,
where the ordered state breaks $U(1)\simeq O(2)$ symmetry.
The chiral $O(3)$-GNY model describes the transition towards
an antiferromagnetic order that breaks $SU(2)$ symmetry in related materials.
These are three typical examples in the conventional chiral $O(N)$ universality class.
For the chiral emergent-$O(N)$ universality class,
however, it cannot be achieved by breaking a single conventional symemtry.
In fact, the emergent symmetry can be realized in some special cases when two ordered phases
with different broken symmetry meet at a common critical point\cite{janssen2,roy3}.

Limited by the constraint on $N$,
the emergent-$O(N)$ symmetry depends on the number of fermion flavor and has an upper boundary $O(2N_f+3)$.
Specifically, in the Dirac system with single flavor ($N_f=1$) of four-component fermions,
the emergent-$O(4)$ and $O(5)$ symmetries at criticality are possible.
And in the system with two flavors ($N_f=2$) of four-component fermions,
the emergent-$O(N)$ symmetry for $N=4, 5, 6, 7$ are possible to be found.
More importantly, since the constraint on $N$ are obtained from the leading-order anomalous dimensions which
are not affected by higher-order corrections,
thus the constraint holds in any higher-order computations.
What we want to emphasize in particular is that close enough to the critical point the emergent symmetry-$O(2)$ ($Z_2\times Z_2$) and $O(3)$ ($Z_2\times O(2)$\cite{roy3}) are also compatible with the constraint.

Finally, let us briefly discuss the emergent supersymmetric critical point.
For $N_f=1/4$, the GNY model in the chiral Ising universality class
exhibits an emergent supersymmetry,
the YWF2 fixed point now becomes a supersymmetric fixed point at which $\eta_{\phi}=\eta_{\psi}=\epsilon/7$.
Another interesting case is the $N_f=1/2$ GNY model in the chiral XY universality class\cite{lzx3},
this version of model has been argued that the supersymmetry might emerges at the critical point,
with the anomalous dimensions are given by
$\eta_{\phi}=\eta_{\psi}=\epsilon/3$.
In general, we note from Eq.(\ref{abb}) that the supersymmetry is ensured by $N=4N_f$.
As a result, the supersymmetry is expected to be emerged from the quantum
critical point characterized by an emergent-$O(4)$ symmetry,
such supersymmetric critical point is expected to be found
in the chiral GNY model with single flavor ($N_f=1$) of four-component fermions.

\begin{table}[!tb]
\centering
\caption{ Numerical series of the inverse correlation length exponent $\nu_y^{-1}$ at the stable YWF2 fixed point.}
\resizebox{\linewidth}{!}
{
\begin{tabular}{lll}
\hline\hline
Basic $O(N)$ symmetry\hspace{0.5cm} &$N_f=1$\hspace{1.0cm} &$N_f=2$\\
\hline
$Z_2$ ($N$=1)               &$2-0.8347\epsilon$    &$2-0.9524\epsilon$ \\
$O(2)$                      &$2-1.1325\epsilon$    &$2-1.2\epsilon$ \\
$O(3)$                      &$2-1.5988\epsilon$    &$2-1.5273\epsilon$ \\
\\
Emergent $O(N)$ symmetry \hspace{0.5cm} &$N_f=1$\hspace{1.0cm} &$N_f=2$\\
$O(4)$                      &$2-2.5\epsilon$       &$2-2.0\epsilon$ \\
$O(5)$                      &$2-5.1583\epsilon$    &$2-2.7692\epsilon$ \\
$O(6)$                      &not exist             &$2-4.2857\epsilon$ \\
$O(7)$                      &not exist             &$2-8.8\epsilon$ \\
\hline\hline
\end{tabular}
}
\label{tab1}
\end{table}

\section{Conclusions and comments} \label{section5}
Within the first-order $\epsilon$ expansion, we have studied the critical structure and the emergent symmetry of chiral GNY model
in the presence of a small $O(N)$-anisotropy.
This model includes a Yukawa term with $N_f$ flavors of four-component Dirac fermions strongly
coupled to an $O(N)$ scalar field.
We have determined the stability of the fixed points and computed the critical exponents
by means of perturbative renormalization in $4-\epsilon$ dimensions.
On the basis of $O(N)$-GNY model, we have discussed the physically reasonable emergent symmetry in Dirac systems.
Further, the supersymmetric quantum critical point in the emergent-$O(4)$ universality class have also been discussed briefly.
The main three conclusions of our findings can be summarized as follows.

(i) The GNY model in the chiral Ising, chiral XY, or chiral $O(3)$ universality class has an unique infrared-stable fixed point,
the so called Wilson-Fisher-Yukawa fixed point.
For the GNY model in the chiral emergent-$O(N)$ universality class with $N\geq4$, in order to
meet the requirements of measurability for Yukawa coupling,
the chiral emergent-$O(N)$ universality class is physically meaningful if and only if $N$ is less than $2N_f+4$,
where $N_f$ is the number of flavors of four-component Dirac fermions.
On the premise that the emergent-$O(N)$ universality class is meaningful,
the GNY model in the chiral emergent-$O(N)$ universality class is also dominated by the Wilson-Fisher-Yukawa fixed point.
This result holds at least in the cases with $N_f=1$ and $N_f=2$.
As a result, the small $O(N)$-anisotropy that breaks $O(N)$ symmetry is irrelevant in the chiral emergent-$O(N)$ universality class.

(ii) The non-negativity of the anomalous dimensions and the measurability of the Yukawa coupling
impose the constraint $N<2N_f+4$ on the emergent-$O(N)$ symmetry.
As a result, the emergent symmetry has an upper boundary $O(2N_f+3)$ in Dirac systems.
The enlarged emergent-$O(4)$ and $O(5)$ symmetries are possible to be found in the system with single flavor ($N_f=1$) of four-component fermions,
and the enlarged emergent-$O(4)$, $O(5)$, $O(6)$ and $O(7)$ symmetries are expected to be
found in the systems with two flavors ($N_f=2$) of four-component fermions.
These results hold in any higher-loops calculations.
Moreover, the emergent-$O(2)$ ($Z_2\times Z_2$) and $O(3)$ ($Z_2\times O(2)$\cite{roy3})
symmetries are also compatible with the constraint.

(iii) In the chiral emergent-$O(4)$ universality class, there is a supersymmetric critical point,
with the anomalous dimensions $\eta_{\phi}=\eta_{\psi}=\epsilon$.
The supersymmetry is expected to be found in the system with fermion flavor $N_f=1$.

Our result has close connections with the recent study
on the quantum multicritical point that possesses enlarged emergent-$O(N)$ symmetry in Dirac systems.
Refs.[\onlinecite{janssen2}] and [\onlinecite{roy3}] pointed that the multicritical point
between the distinct $O(s_1)$ and $O(s_2)$ symmetry broken phases is generically characterized by an emergent
$O(s_1+s_2)$ symmetry. For the conventional broken symmetries in Dirac systems, e.g., $Z_2$, $O(2)$ and $O(3)$,
our result agrees well with Refs.[\onlinecite{janssen2}] and [\onlinecite{roy3}].
In addition, our result also suggests some rich emergent symmetries
for fermionic criticality in graphene-like systems, for instance,
the deconfined transition with emergent $Z_2\times Z_2\times O(2)$,
$Z_2\times O(2)\times O(2)$ and $O(2)\times O(2)\times O(3)$ symmetries and so on.
This conjecture deserves further investigation in future.
Importantly, our result is appliable to the recently observed emergent-$O(4)$ symmetry\cite{sato1},
the deconfined transition between $SO(3)$-semimetal and $U(1)$-insulator\cite{lzh1}, or the transition between antiferromagnetism and valence-bond-solid in quantum Monte Calor simulations of a designed synthetic Dirac systems\cite{lzx5}.

Finally, the new emergent-$O(4)$ supersymmetric critical point is interesting on its own.
In the future, we expect such supersymmetric critical point could be cross-checked by other methods,
i.e., conformal bootstrap approach and quantum Monte Carlo simulation\cite{lzx3}.

\begin{acknowledgments}
We acknowledge the support from the startup grant under No.20175788 in Guizhou University.
\end{acknowledgments}

\begin{appendix}\label{appdendixa}
\section{Renormalization constants at one-loop order}\label{appendixa}
This appendix devotes to calculate the renormalization constants.
At one-loop order, the renormalization constants are defined as $Z_X=1+\delta_X$,
these $\delta_X$s are known as counterterms which are used to absorb the divergencies in the 1PI diagram.
Therefore, we need to calculate all the divergence of the 1PI diagrams in Fig. \ref{fig1}.
\subsection{Vertex tensor product}
Before the calculation, let us derive the vertex product contributing to the effective quartic boson interaction.
The general vertex tensor can be represented as the following symmetrized form:
$\lambda_{ijkl}=\lambda_1T^{(1)}_{ijkl}+\lambda_2T^{(2)}_{ijkl}$, with
\begin{align}
    T^{(1)}_{ijkl}=\frac13(\delta_{ij}\delta_{kl}+\delta_{ik}\delta_{kl}+\delta_{il}\delta_{jk}),
\end{align}
\begin{align}
    T^{(2)}_{ijkl}=\delta_{ijkl}=\left\{
    \begin{array}{cc}
     1,& i=j=k=l, \\
     0,& \text{otherwise}.
    \end{array}
    \right.
\end{align}
Here, $\delta_{ijkl}$ satisfies $\sum_k\delta_{ijkk}=\delta_{ij}$, $\sum_i\delta_{ii}=N$.
The leading order 1PI diagram for quartic boson interaction need to calculate the tensor product: $\lambda_{ijmn}\lambda_{mnkl}$.
For $T^{(1)}T^{(1)}$, we have
\begin{equation}
    T^{(1)}_{ijmn}T^{(1)}_{mnkl}=\frac19[(N+4)\delta_{ij}\delta_{kl}+2\delta_{ik}\delta_{jl}+2\delta_{il}\delta_{jk}].
\end{equation}
After symmetrization, we have the symmetrized result
\begin{align}
    \left[T^{(1)}_{ijmn}T^{(1)}_{mnkl}\right]_s=\frac{N+8}{9}T^{(1)}_{ijkl}.
\end{align}
Accordingly,
\begin{gather}
    \left[T^{(1)}_{ijmn}T^{(2)}_{mnkl}+T^{(2)}_{ijmn}T^{(1)}_{mnkl}\right]_s=\frac{2}{3}T^{(1)}_{ijkl}+\frac{4}{3}T^{(2)}_{ijkl},\\
    \left[T^{(2)}_{ijmn}T^{(2)}_{mnkl}\right]_s=T^{(2)}_{ijkl}.
\end{gather}
Combining the above results together, the symmetrized tensor product is given by
\begin{align}\label{stp}
   \left[\lambda_{ijmn}\lambda_{mnkl}\right]_s=&\left[\frac{N+8}{9}\lambda^2_1+ \frac{2\lambda_1\lambda_2}{3}\right]T^{(1)}_{ijkl}\notag\\
                                +&\left[\lambda^2_2+ \frac{4\lambda_1\lambda_2}{3}\right]T^{(2)}_{ijkl}.
\end{align}
Finally, the product $T^{(1)}_{ijmn}\delta_{mn}=(N+2)\delta_{ij}/3$ will be necessary for boson mass-squared renormalization factor.

\subsection{Boson two-point function}
The 1PI boson two-point function are given in Fig.\ref{fig1}(a)-(c). Fig.\ref{fig1}(a) contributes to $Z_{\phi}$,
(b) and (c) contribute to the mass-square renormalization $Z_{m^2}$. Fig.\ref{fig1}(a) gives
\begin{align}
\mathcal{A}_{1a}(p)&=(-1)(-g)^{2}N_{f}\int \frac{d^dk}{(2\pi)^d}
\text{Tr}\left[\Sigma_a\frac{1}{i\slashed{k}}\Sigma_{b}\frac{1}{i(\slashed{k}+\slashed{p})}\right] \notag \\
&=g^2N_fD_{\Sigma}\delta_{ab}\int \frac{d^dk}{(2\pi)^d}\left[ \frac{1}{\slashed{k}}\frac{1}{(\slashed{k}+\slashed{p})}\right].
\end{align}
where the minus sign $(-1)$ arises from fermion loop. The integral is a standard Feynman integral, introduce a
Feynman parameter and perform the elementary integral, then we have
\begin{align}
\mathcal{A}_{1a}(p)=-g^2N_fD_{\Sigma}\frac{K}{\epsilon}\delta_{ab}p^2,
\end{align}
where $D_{\Sigma}$ is the dimensions of $\Sigma_a$, $K=1/(4\pi)^2$ here and after in this paper.
To reach the final result, the commutating rule $[\Sigma_a,\gamma_{\mu}]=0$ has been used [see Eq.(\ref{comm})].
We only extract the divergences when $d\rightarrow 4$
since only the divergent part should cancel out with the $Z$-factors, that is, these $Z$-factors depend only on the divergences in the dimensional regularization and MS scheme.

For Fig.\ref{fig1}(b) and (c), we have
\begin{align}
\mathcal{A}_{1b}+\mathcal{A}_{1c}&=\frac{1}{2}(-\lambda_1T^{(1)}_{ijkl}-\lambda_2T^{(2)}_{ijkl})
\int \frac{d^{d}k}{(2\pi)^d}\frac{i\delta ^{kl}}{k^2+m^2} \notag\\
&=\left[\frac{(N+2)}{3}\lambda_1+\lambda_2\right]m^2\frac{K}{\epsilon }\delta^{ij}.
\end{align}
To reach the final result, we have used the useful integral in d-dimensional Euclidean space:
\begin{equation}\label{fi}
 \int\frac{d^dk}{(2\pi)^d}\frac1{(k^2_E+\Delta)^n}=\frac1{(4\pi)^{d/2}}\frac{\Gamma(2-d/2)}{\Gamma(n)}\frac1{\Delta^{2-d/2}}.
\end{equation}
In terms of the renormalization condition
\begin{equation}
    \mathcal{A}_{1a}+\mathcal{A}_{1b}+\mathcal{A}_{1c}-p^2\delta_{\phi}-(\delta_{\phi}+\delta_{m^2})m^2=0,
\end{equation}
we find
\begin{align}
 Z_{\phi}&=1-g^2N_fD_{\Sigma}\frac{K}{\epsilon},\\
 Z_{m^2}&=1+\left[\frac{(N+2)}{3}\lambda_1+\lambda_2+g^2N_fD_{\Sigma}\right]\frac{K}{\epsilon}.
\end{align}

\subsection{Fermion two-point function}
Fig.\ref{fig1}(d) shows the 1PI fermion two-point function, which is given by
\begin{align*}
\mathcal{A}_{1d}(p)=(-g)^{2}\int \frac{d^dk}{(2\pi)^d}
\Sigma_a\frac{-i\slashed{k}}{k^2}\Sigma_b\frac{\delta_{ab}}{(p-k)^2+m^2}.
\end{align*}
Using the feynman parameter integral:
\begin{equation*}
    \frac1{k^2}\frac1{(p-k)^2+m^2}=\int^1_0dx\frac1{[(k+p(x-1))^2+\Delta]^2},
\end{equation*}
where $\Delta=p^2x(1-x)+m^2(1-x)$, then shifting the integration variable, $k\rightarrow k-p(x-1)$,
applying the result in Eq.(\ref{fi}), one obtains
\begin{align}
    \mathcal{A}_{1d}&=g^2N(-i)\int^1_0dx\int\frac{d^dk}{(2\pi)^d}\frac{-\slashed{p}(x-1)}{(k^2+\Delta)^2},\notag\\
                    &=-g^2N\frac{K}{\epsilon}i\slashed{p}.
\end{align}
The divergence in Fig.\ref{fig1}(d) should cancel out with $\delta_{\psi}$, the renormalization condition is
$-i\slashed{p}\delta_{\psi}+\mathcal{A}_{1d}(p)=0$, we thus obtain
\begin{equation}
    Z_{\psi}=1-g^2N\frac{K}{\epsilon}.
\end{equation}

\subsection{Yukawa vertex}
To the one-loop, there is a single diagram for the 1PI Yukawa vortex, see Fig.\ref{fig1}(e).
It can be calculated to give
\begin{align*}
\mathcal{A}_{1d}(p,q)&=\int \frac{d^{d}k}{(2\pi)^{d}}\frac{\delta _{ac}}{
(p-k)^{2}+m^{2}}\times\\
&\left[(-g\Sigma_{a})\frac{1}{i(\slashed{q}+\slashed{k})}(-g\Sigma_{b})\frac{1}{i\slashed{k}}(-g\Sigma_{c})\right].
\end{align*}
The divergent part can be obtained by setting the external momentum $p$, $q$ to be zero.
Making use of the general Feynman parameters:
\begin{equation}
\frac{1}{AB^{n}}=\int_{0}^{1}dxdy\delta (x+y-1)\frac{ny^{n-1}}{(xA+yB)^{n+1}},
\end{equation}
we have
\begin{align}
\mathcal{A}_{1d}(0,0)&=g^{3}\Sigma_{b}(2-N)\int \frac{d^{d}k}{(2\pi )^{d}}\frac{1}{k^{2}+m^{2}}\frac{1}{k^{2}} \notag\\
&=g^{3}(2-N)\Sigma_{b}\int_{0}^{1}dx\int\frac{d^{d}k}{(2\pi)^{d}}\frac{1}{(k^{2}+xm^{2})^{2}} \notag\\
&=g^{3}2(2-N)\Sigma_{b}\frac{K}{\epsilon}.
\end{align}
The divergences should cancel out with $Z_gZ_{\psi}\sqrt{Z_{\phi}}-1$, so, it is sufficient to define the renormalization condition
\begin{equation}
  -g\Sigma_b(Z_gZ_{\psi}\sqrt{Z_{\phi}}-1)+\mathcal{A}_{1d}(0,0)=0.
\end{equation}
Inserting $Z_{\psi}, Z_{\phi}$ into it leads to
\begin{equation}
    Z_{g}=1+g^2[(4-N+N_fD_{\Sigma}/2)]\frac{K}{\epsilon}.
\end{equation}

\subsection{$O(N)$-symmetric bosonic quartic-vertex}
The relevant diagrams for $O(N)$ bosonic quartic-vertex are given in Fig.\ref{fig1}(f)-(h).
Fig.\ref{fig1}(f) and (g) contribute partly to the boson self-interaction $\lambda_1$ [see Eq.(\ref{stp})], the result can be evaluated to give
\begin{align}
  \mathcal{A}^{(1)}_{1f}+\mathcal{A}^{(1)}_{1g}&=\frac32[\lambda_{ijmn}\lambda_{mnkl}]^{(1)}_s
  \int\frac{d^dk}{(2\pi)^d}\left[\frac{1}{k^2+m^2}\right]^2\notag\\
  &=3\left[\frac{N+8}{9}\lambda^2_1+ \frac{2\lambda_1\lambda_2}{3}\right]T^{(1)}_{ijkl}\frac{K}{\epsilon},
\end{align}
where the number $3$ counts $s$-channel, $t$-channel and $u$-channel in all. We have set all external momentum to be zero to get the divergences.

For Fig.\ref{fig1}(h) with the the external momentum $p=0$, we have
\begin{align}
\mathcal{A}^{(1)}_{1h}(0)&=-2N_fg^4\int \frac{d^dk}{(2\pi)^d}
{\text{Tr}\left[\Sigma_i\frac1{i\slashed{k}}\Sigma_j\frac1{i\slashed{k}}\Sigma_k\frac1{i\slashed{k}}
\Sigma_l\frac1{i\slashed{k}}\right]}\notag\\
&=-2N_fg^4\int\frac{d^dk}{(2\pi)^d}\frac1{k^4}\text{Tr}[\Sigma_i\Sigma_j\Sigma_k\Sigma_k],
\end{align}
where Tr[...] in the first line denotes the trace over the sigma matrix,
the number $2$ in the first line counts the exchange of $i$ and $j$ (or equivalently, $k$ and $l$).
Using the identity
\begin{equation}
    \text{Tr}[\Sigma_i\Sigma_j\Sigma_k\Sigma_k]=D_{\Sigma}(\delta_{ij}\delta_{kl}-\delta_{ik}\delta_{jl}+\delta_{il}\delta_{jl}),
\end{equation}
and performing the $k$ integration, we obtain
\begin{align}
\mathcal{A}^{(1)}_{1h}(0)=-12N_fD_{\Sigma}g^4\frac{K}{\epsilon}T^{(1)}_{ijkl}.
\end{align}
To extract the renormalization constant $ Z_{\lambda_1}$,
we define the renormalization condition
\begin{equation}
    -\lambda_1(Z_{\lambda_1}Z^2_{\phi}-1)T^{(1)}_{ijkl}+\mathcal{A}^{(1)}_{1f}+\mathcal{A}^{(1)}_{1g}+\mathcal{A}^{(1)}_{1h}(0)=0,
\end{equation}
Inserting $Z_{\phi}$ into this condition leads to
\begin{align}
    Z_{\lambda_1}&=1+[\frac{N+8}{3}\lambda_1+2\lambda_2-12D_{\Sigma}N_fg^4/\lambda_1\notag \\
    &+2N_fD_{\Sigma}g^2]\frac{K}{\epsilon}.
\end{align}

\subsection{$O(N)$-anisotropy}
The diagrams for $O(N)$-anisotropy are given in Fig.\ref{fig1}(i) and (j),
and the divergence can be computed to give
\begin{align}
  \mathcal{A}^{(2)}_{1i}+\mathcal{A}^{(2)}_{1j}&=\frac32[\lambda_{ijmn}\lambda_{mnkl}]^{(2)}_s
  \int\frac{d^dk}{(2\pi)^d}\left[\frac{1}{k^2+m^2}\right]^2\notag\\
  &=3\left[\lambda^2_2+ \frac{4\lambda_1\lambda_2}{3}\right]T^{(2)}_{ijkl}\frac{K}{\epsilon}.
\end{align}
Here, to reach the final result, we have used the symmetrized tensor product Eq.(\ref{stp}).
Defining the renormalization condition
\begin{equation}
    -\lambda_2(Z_{\lambda_2}Z^2_{\phi}-1)T^{(2)}_{ijkl}+\mathcal{A}^{(2)}_{1i}+\mathcal{A}^{(2)}_{1j}=0,
\end{equation}
and inserting $Z_{\phi}$ into it, we have
\begin{align}
    Z_{\lambda_2}&=1+[3\lambda_2+4\lambda_1+2N_fD_{\Sigma}g^2]\frac{K}{\epsilon}.
\end{align}
Making use of the sigma matrix with dimensions four but without asking their explicit representation,
we obtain all the renormalization constants of the main text.

\end{appendix}

\end{document}